\documentclass[twoside]{article}
\usepackage{fleqn,espcrc2}

\usepackage{graphicx}
\usepackage[figuresright]{rotating}

\newcommand{\AmS}{{\protect\the\textfont2
  A\kern-.1667em\lower.5ex\hbox{M}\kern-.125emS}}

\title{Domain walls in supersymmetric QCD\thanks{Work partly
     supported by the British Council/Acciones Integradas programme 
     under contract HB1997-0073.}}

\author{B.~de~Carlos\address[SUSX]{Centre for Theoretical Physics, 
        University of Sussex, \\ 
        Falmer, Brighton BN1 9QJ, United Kingdom},
        M.B.~Hindmarsh\addressmark[SUSX],
        N.~McNair\addressmark[SUSX]\thanks{Work supported by PPARC.}
         and
        J.M.~Moreno\address{Instituto de Estructura de la Materia,
        CSIC \\ Serrano 123, 28006 Madrid, Spain}\thanks{Work
        supported by CICYT of Spain under contract AEN98-0816, and by
     the EU under contract HPRN-CT-2000-00152.}}
       
\begin{document}

\begin{abstract}
In this talk we construct BPS-saturated domain walls in supersymmetric 
QCD, for any values of the masses of the chiral matter superfields.
We compare our results to those already obtained in the literature and
we also discuss their range of applicability, as well as future 
directions that would be desirable to explore in order to achieve a
complete understanding of supersymmetric gluodynamics as a step in
improving our knowledge of how QCD works.
\vspace{1pc}
\end{abstract}


\maketitle

\leftline{SUSX-TH/01-009}
\leftline{IEM-FT-210/01} 
\leftline{IFT-UAM/CSIC-01-05}

\section{Introduction}

Supersymmetric (SUSY) gluodynamics, the theory of gluons and 
gluinos with gauge group SU($N_c$), shares a lot of common features
with ordinary QCD, with the advantage of being supersymmetric. 
Therefore it seems a good idea to exploit the supersymmetric properties 
of SUSY gluodynamics in order to try to understand better how QCD 
works. Although this is an idea which has been around for many
years, the recent and positive results of Dvali and 
Shifman~\cite{Dvali97}, which we shall briefly explain in detail, 
triggered a great deal of activity in these past three years.
     
Our starting point will be the Lagrangian for SUSY gluodynamics, 
which is given by 
\begin{equation}
{\cal L} = \frac{1}{g_0^2} \left[ \frac{1}{4} {\cal G}_{\mu\nu}^a
{\cal G}_{\mu\nu}^a + i \lambda^{a \alpha} 
{\cal D}_{\alpha \dot{\beta}} \bar{\lambda}^{a \dot{\beta}} \right] \;\;,
\label{lag}
\end{equation}
where $g_0$ is the gauge coupling constant, $G_{\mu\nu}^a$ the usual
gluon field strength tensor and $\lambda$ is the gaugino field. This 
theory has an axial U(1) symmetry which is broken down to $Z_{2N_c}$, 
the chiral symmetry, by the anomaly. Moreover, the theory is strongly 
coupled: at a scale $\Lambda$ gaugino condensates (which, from now
on, we shall denote as $\langle \lambda \lambda \rangle$) will form.
This will break the chiral symmetry from $Z_{2N_c}$ to $Z_2$ and, 
clearly, the gaugino condensate is the order parameter associated with
the symmetry breakdown.
     
The value of the gaugino condensate was calculated by various methods 
a few years ago~\cite{Witte82}. The $N_c$ degenerate vacua of our 
theory are given by
\begin{equation}
\langle \lambda \lambda \rangle \equiv \langle {\rm Tr} \lambda^a 
\lambda_a \rangle = \Lambda^3 e^{i 2 \pi \frac{k}{N_c}}, k=0, 
\ldots, N_c-1.
\label{cond}
\end{equation}
The presence of degenerate vacua leads immediately to the formation
of domain walls interpolating between them. The key result of Dvali
and Shifman~\cite{Dvali97} was to realize that if these walls are of 
a certain kind, i.e. BPS-saturated, then they preserve one half of 
the supersymmetry and their energy density is {\em exactly}
calculable and given by
\begin{equation}
\epsilon = \frac{N_c}{8 \pi^2} \left| \langle {\rm Tr} \lambda^a 
\lambda_a \rangle_{\infty} -  \langle {\rm Tr} \lambda^a \lambda_a 
\rangle_{-\infty}   \right| \;, 
\label{enden}
\end{equation}
i.e. the difference between the values of the condensate at the two
vacua -situated at spatial infinities- between which the wall is 
interpolating. This is very interesting because it has been 
suggested~\cite{Witte97} that BPS-saturated domain walls would play
an important role in the D-brane description of N=1 supersymmetric
QCD (SQCD). Also, recent work claims that, in the large $N_c$ limit,
these walls are BPS-saturated states~\cite{Dvali99}, but so far it 
has not been fully proved for finite $N_c$ configurations. In any case
the question to be answered, before making any further connections to 
other interesting topics, is what the nature of these domain walls is. 
     
In order to do that we are going to discuss first of all different 
effective approaches to SUSY gluodynamics and the problems associated 
with them. Then we shall study in detail the Taylor, Veneziano, 
Yankielowicz (TVY) approach~\cite{Taylo83} which is the one used in 
our calculation. Finally we present results and draw a few conclusions. 
     
\section{Towards an effective theory of SUSY gluodynamics}    
   
A first and very complete description of SUSY gluodynamics was 
provided by Veneziano and Yankielowicz~\cite{Venez82} (VY), in terms 
of a composite chiral superfield $S$ whose lowest component is the
gaugino condensate
\begin{equation}
S \equiv \frac{3}{32 \pi^2} \langle {\rm Tr} (\omega_{\alpha} 
\omega^{\alpha}) \rangle  = \frac{3}{32 \pi^2}  \langle {\rm Tr} 
(\lambda_{\alpha} \lambda^{\alpha}) \rangle + \ldots
\label{comp}
\end{equation}
The Lagrangian for this model is given by the usual supersymmetric 
structure,
\begin{equation}
{\cal L} = \frac{1}{4} \int d^4 \theta {\cal K} + \frac{1}{2}
\left[ \int d^2 \theta {\cal W} + {\rm h.c.} \right] \;\;,
\label{lag_susy}
\end{equation}
with ${\cal K}$ the K\"ahler potential and ${\cal W}$ the
superpotential. Throughout this talk we shall be working with
canonically normalized fields, therefore ${\cal K}=(\bar{S} S)^{1/3}$.
The superpotential for the VY model is given by
\begin{equation}
{\cal W} = \frac{2}{3} S \ln \left( \frac{S^{N_c}}{(e\Lambda^3)^{N_c}} 
\right) \;\;.
\label{supTV}
\end{equation}
The structure of ${\cal W}$ is uniquely fixed by the anomaly and the 
symmetries of the theory. Again $\Lambda$ is the scale parameter. It
has been widely shown that the VY superpotential describes very well
the vacuum structure of SUSY gluodynamics, however this is not the
case when we try to describe the dynamics of the theory and, in
particular, when we try to construct domain walls.
     
It was pointed out not so long ago~\cite{Kovne97} that this Lagrangian
is not explicitly $Z_{2N_c}$ invariant. Moreover, the scalar potential
derived in global supersymmetry, i.e. $V(S) = {\cal K}_{S\bar{S}}^{-1} 
|\partial {\cal W}/ \partial S|^2$ is multivalued due to the presence 
of the logarithm in ${\cal W}$. That is, a physical state and its 
equivalent, just rotated by $2 \pi$, would give rise to completely 
different values of $V$, which is totally unacceptable.
     
In order to cure this problem, several solutions have been proposed.
In Refs~\cite{Kovne97,Kogan98} the idea of {\em glued potential} was
introduced and developed. Essentially it consists of adding a Lagrange
multiplier to the Lagrangian in such a way that the scalar potential
is divided into $N_c$ sectors which, glued together, result in a 
single-valued $Z_{2N_c}$-invariant theory. The problem with such an
amendment is that {\em cusps} will inevitably form at the joining
point of each sector with its nearest neighbours. Any configuration
interpolating between two vacua will necessarily cross a cusp and it
is doubtful whether it would be possible to correctly interpret the
energy density associated with it. Therefore we conclude that the VY
model, even when properly modified, is unable to provide us with a 
good description of domain walls in SUSY gluodynamics.
     
Other solutions have been proposed to deal with the problems
associated to the presence of the logarithm in the VY superpotential.
For example, in Ref.~\cite{Dvali99} it was proposed that the origin
of such problems was related to leaving behind some relevant degrees 
of freedom when deriving the VY efective Lagrangian. It is then 
suggested that introducing new degrees of freedom in the model, in 
particular a glueball order parameter, would be enough to enable us
to construct well behaved domain walls. This is certainly a
possibility worth considering, however we will not be following this 
approach but that of Taylor, Veneziano and 
Yankielowicz~\cite{Taylo83} (TVY), which we describe in detail in the 
next section.

\section{Our model: the TVY approach}
     
Given all the problems of the VY model when trying to construct domain
wall solutions of the equations of motion, it is convenient to work 
with a slightly more complicated model which respects all the symmetries
of the theory. This consists of adding $N_f$ pairs of chiral 
superfields, $Q^i$, $\bar{Q}_i$ to the VY model. Then, below the
condensation scale we shall have matter condensates, $M_j^i = Q^i 
\bar{Q}_j$, as well as gaugino condensates. The Lagrangian, given
again by Eq.~(\ref{lag_susy}), has the following superpotential
\begin{equation}
{\cal W}  =  \frac{2}{3} S \ln \left( \frac{S^{N_c-N_f} {\rm det} M}
{e^{N_c-N_f} \Lambda^{3N_c-N_f}} \right) - \frac{1}{2} {\rm Tr}(mM).
\label{TVY}
\end{equation}
From now on we shall work in a flavour-diagonal basis, i.e. $m_k^j =
\delta_k^j m_j$. Our goal will be to construct domain walls between the 
vacua of this model, and then try to obtain the limit of large masses 
where the theory should tend to SUSY gluodynamics. In other words, we 
start off by working in the weak (Higgs) phase of the model, and we
will try to extrapolate the results obtained to strong coupling.

Therefore the first step consists of defining the vacua of the 
model. These are given by
\begin{eqnarray}
S_*^{N_c} & = & \left(\frac{3}{4}\right)^{N_f} \det m \nonumber \\
     & & \label{minima}  \\
(M_i^j)_* & = & \delta_i^j \frac{1}{m_i} \frac{4}{3} S_* \;\;, \nonumber 
\end{eqnarray}
where, and from now on, we are taking $\Lambda=1$. That is, the vacuum 
values for the gaugino and matter condensates are aligned ($*$ denotes 
values at the vacuum). Also, one can easily evaluate the
superpotential to find out that it is proportional to the vacuum 
values of the fields, i.e. $W_* = -(2/3) N_c S_*$.
     
A very important point which we are going to discuss in detail is 
the paths the fields take when going from one vacuum to another.
As mentioned before, we are using the matter fields to restore 
the lost $Z_{2N_c}$ invariance of the TV Lagrangian and ensure that,
when building domain walls, we do not cross the logarithmic branch.
In order to formulate this in a more precise way, let us suppose that
we are building a wall between vacuum {\em a} and vacuum {\em b}. 
Then we define the field trajectories as follows
\begin{eqnarray}
S|_b & = & e^{i \delta} S|_a  \label{paths} \\
M_i^i|_b & = & e^{i(\delta-2 \pi \omega_i)} M_i^i|_a \;,
i=1,\ldots,N_f. \nonumber
\end{eqnarray}
where $\omega_i$, the {\em windings} of the matter fields, have to be
integers, in order to fulfill the alignment condition at the vacua.
To ensure that the logarithmic branch is not crossed one must impose
the following condition, coming from the fact that the phases in the
logarithm of Eq.~(\ref{TVY}) cancel out
\begin{equation}
(N_c-N_f) \delta + \sum_{i=1}^{N_f} (\delta - 2\pi \omega_i) = 0 \;.
\label{phases}
\end{equation}
As an immediate result it follows that $\delta=2 \pi k/N_c$ where we 
define $k=\sum_{i=1}^{N_f} \omega_i$. Note that, if all the windings 
are equal to one, then $k=N_f$. This is an important fact to which we
will return.

Let us now briefly discuss the method used to calculate the 
domain wall profiles. We assume that the walls spread along the
$xy$ plane, therefore the profiles are calculated along $z$. This
is done by minimizing the energy functional which, for a generic 
set of fields, $X^k$, with K\"ahler metric 
$g_i^j={\cal K}_{X^i \bar{X}_j}$, looks like
\begin{equation}
\epsilon_{ab} = \frac{1}{2} \int_{-\infty}^{+\infty} dz (g^j_i 
\partial_z X^i \partial_z \bar{X}_j + (g_i^j)^{-1} {\cal W}_i 
\bar{{\cal W}}^j) \;,
\label{en_fun}
\end{equation}
where ${\cal W}_i = \partial {\cal W}/\partial{X^i}$. 
This equation can be rewritten as
\begin{eqnarray}
\epsilon_{ab} & = & {\rm Re} (e^{i\gamma} ({\cal W}_b-{\cal W}_a)) 
\nonumber \\     
& + &  \frac{1}{2} \int_{-\infty}^{+\infty} dz (g_i^j)^{-1} 
(g_k^j \partial_z X^k - e^{i\gamma} \bar{{\cal W}}^j) \nonumber \\
& \times &  (g_i^k \partial_z \bar{X}_k - e^{-i\gamma} 
{\cal W}_i) \label{en_fun_2} 
\end{eqnarray}    
where $e^{i\gamma}$ is an arbitrary phase. In fact, if we choose it
in a clever enough way, i.e. $e^{-i\gamma_{ab}}=({\cal W}_b-{\cal W}_a)
/|{\cal W}_b-{\cal W}_a|$, then it is easy to see that we can put a
lower bound on the energy density of the wall that interpolates
between vacua $a$ and $b$, which is
\begin{equation}    
    \epsilon_{ab} \geq |{\cal W}_b-{\cal W}_a| \;\;.
\label{bps}
\end{equation}
This is the so-called BPS bound; if the bound is saturated the wall to
which it corresponds is a BPS-saturated domain wall. An immediate 
consequence of the BPS condition being fulfilled is that the first term
in Eq.~(\ref{en_fun_2}) becomes zero, in other words, the equations
of motion become first order, which represents a significant 
simplification from the numerical point of view. Further details of
this can be found in Ref.~\cite{Chibi97}.
     
In particular, for the TVY model we are working with, the BPS
equations are
\begin{eqnarray}
{\cal K}_{S\bar{S}} \partial_z \bar{S} & = & e^{i\gamma} 
\frac{\partial {\cal W}}{\partial S} \;\;, \nonumber \\
     & & \label{TVY_bps} \\
{\cal K}_{M\bar{M}} \partial_z \bar{M_i^i} & = & e^{i\gamma} 
\frac{\partial {\cal W}}{\partial M_i^i} \;\;. \nonumber  
\end{eqnarray}
In order to simplify the analysis, from now on we will work with
a K\"ahler potential which is canonical for the dimension one matter
fields. For the case of degenerate windings, which we will analyze
next, this is equivalent to taking ${\rm Tr} (M\bar{M})^{1/2}$ modulo
factors of $\sqrt{N_f}$. The case of non-degenerate windings is
slightly more involved and will be fully discussed in the future
\cite{Decar01}. Also we choose $\gamma = -\frac{1}{2} (\delta+\pi) = 
-\frac{k\pi}{N_c} -\frac{\pi}{2}$ where, we stress again, $k$ is the 
sum of all the windings associated with the matter fields. From now
on, in order to find solutions to these equations, we adopt the following
parametrization
\begin{eqnarray}
S(z) & = & |S_*| R(z) e^{i \beta(z)} \nonumber \\
     & & \label{param} \\
M_i^i(z) & = & |M_*| \rho_i(z) e^{i \alpha_i(z)} \;\;, \nonumber
\end{eqnarray}
where $R$, $\beta$, $\rho_i$ and $\alpha_i$  are real functions of the
coordinate $z$. Note that the moduli of the fields, $R$ and $\rho_i$,
are normalized to one at the vacua. The boundary conditions, when 
going from vacuum $j$ to vacuum $j+1$, are
\begin{equation}     
S \rightarrow S e^{i2\pi \frac{k}{N_c}} \;\; , \;\; M_i^i 
\rightarrow M_i^i e^{i2\pi (\frac{k}{N_c}-\omega_i)} \;\;.
\end{equation}     
We also have a constraint, which is a consequence of the equations
of motion, Eqs~(\ref{TVY_bps}), namely
\begin{equation}
{\rm Im}(e^{i\gamma} {\cal W}(S,M_i^i)) = {\rm constant} \;\;,
\label{const}
\end{equation}
which, evaluated at the centre of the domain wall (i.e. $z=0$) becomes
\begin{eqnarray}
& - & R_0  \left[ \left(1-\frac{N_f}{N_c}\right) 
(\ln R_0-1) + \frac{1}{N_c} \sum_i \ln \rho_{0i} \right] 
\nonumber \\
 & + & \frac{1}{N_c} \sum_i (-1)^{\omega_i} \rho_{0i} = \cos \left(
 \pi \frac{k}{N_c} \right) \;\;.
\label{const_TVY}
\end{eqnarray}
This constraint will turn out to be extremely relevant to the 
understanding of the results we are presenting in the next section.
     
\section{Results}

\subsection{Equal windings}
     
To start the discussion of the results, we are going to present those
corresponding to models where the windings $\omega_i$ are all equal
to one (remember that, in such case, $k=N_f$). This is the standard
choice that other authors in the literature have made, and it is
therefore the one we should consider in order to compare with previous
results. In this case the constraint given by Eq.~(\ref{const_TVY})
becomes
\begin{eqnarray}
& - & R_0  \left[ \left(1-\frac{N_f}{N_c}\right) 
(\ln R_0-1) + \frac{N_f}{N_c} \ln \rho_{0} \right] \nonumber  \\     
& - & \frac{N_f}{N_c} \rho_{0} = \cos \left( \pi \frac{N_f}{N_c} 
\right) \;\;.
\label{const_TVY_1}
\end{eqnarray}
Let us summarize the results. The case $N_f=N_c-1$ was analyzed by
Smilga and Veselov in a series of very interesting 
papers~\cite{Smilg97}, and recently reanalyzed by Binosi and ter 
Veldhuis~\cite{Binos00}. They found that the TVY model had 
BPS-saturated domain walls up to a certain value of the mass of the 
chiral matter condensate, $m_*$. Between $m_*$ and another, higher, 
value of the mass, $m_{**}$, they found domain wall solutions which 
were not BPS-saturated. Finally, above $m_{**}$ there were no 
solutions at all. The values of $m_*$, $m_{**}$ seem to depend on 
the value of $N_c$ and they decrease when the latter increases. 
Moreover there seem to be two branches of BPS-saturated solutions 
which merge at $m_*$. These results are very well illustrated by 
Fig.~1 of the third paper in Ref.~\cite{Smilg97}.

\begin{figure}
\centerline{  
\includegraphics[width=7cm]{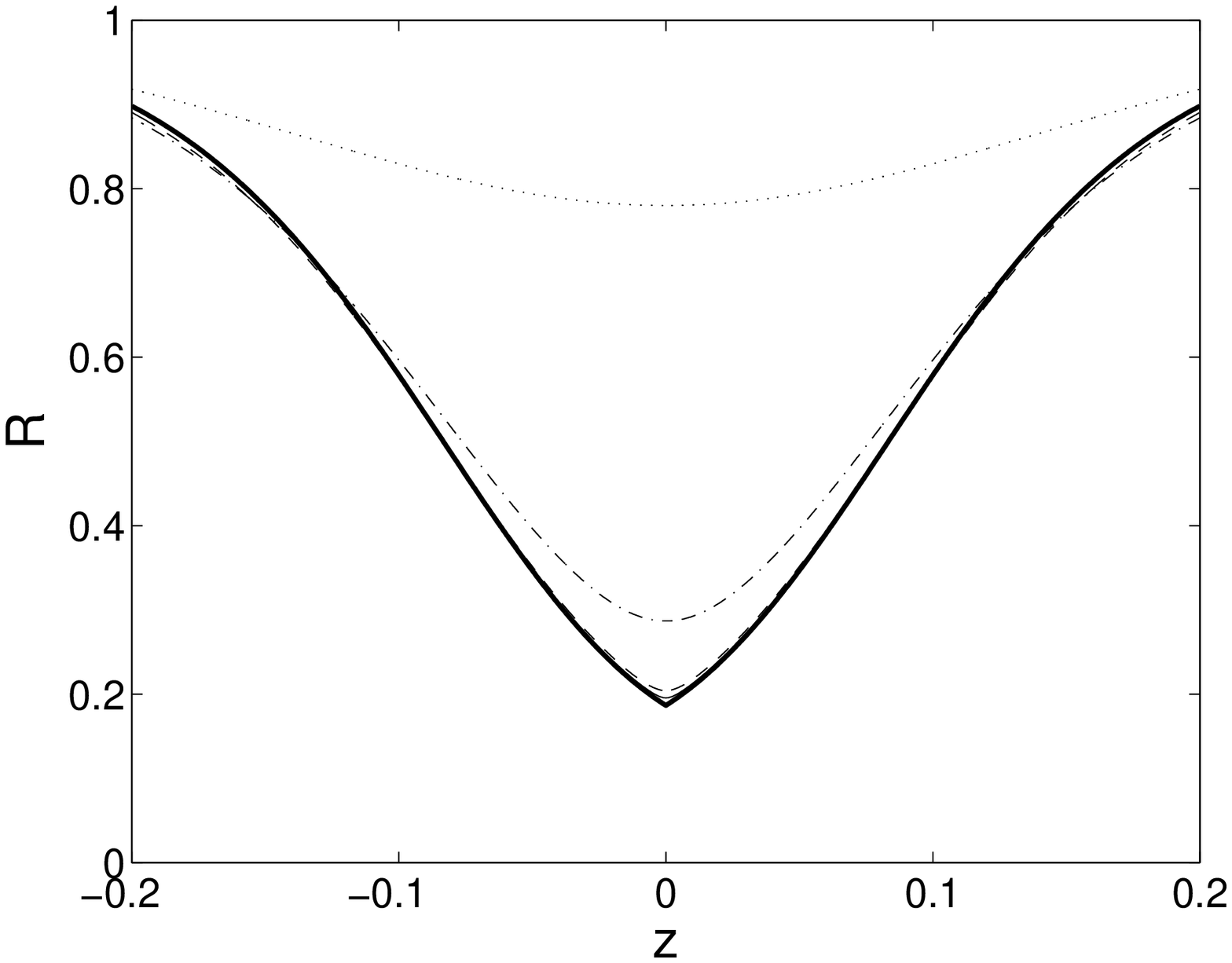}
}
\caption{}
{\footnotesize \noindent \textbf Plot of the magnitude of the
gaugino condensate, $R$, as a function of the spatial coordinate, 
$z$, for $N_c=3$, $N_f=1$ and different values of the mass of the 
matter condensate: $m=2$ (dotted), $m=20$ (dash-dotted), $m=100$ 
(dashed), $m=200$ (solid).
}
\label{fig1}
\end{figure}
A couple of years ago, two of us~\cite{Decar99} analyzed the case of 
$N_f=1$ flavour. We found that there were BPS-saturated solutions for 
any value of the mass parameter $m$. This can be seen in Fig.~1 where we
plot the magnitude of the gaugino condensate, $R$, versus the spatial
coordinate, $z$, for several values of $m$. As the mass increases it can
be seen that the profiles tend to a unique one, represented by the
thick line. We shall return to that point later on. These results 
were subsequently confirmed by the work of Ref.~\cite{Binos00}. In 
general it is possible to make the following statement: 
\begin{itemize}
\item for $N_f/N_c \geq 1/2$ there are BPS-saturated domain walls 
      only up to $m_*$
     
\item for $N_f/N_c < 1/2$ there are BPS-saturated domain walls for
      any value of $m$
\end{itemize}
%
\begin{figure}
\centerline{  
\includegraphics[width=7cm]{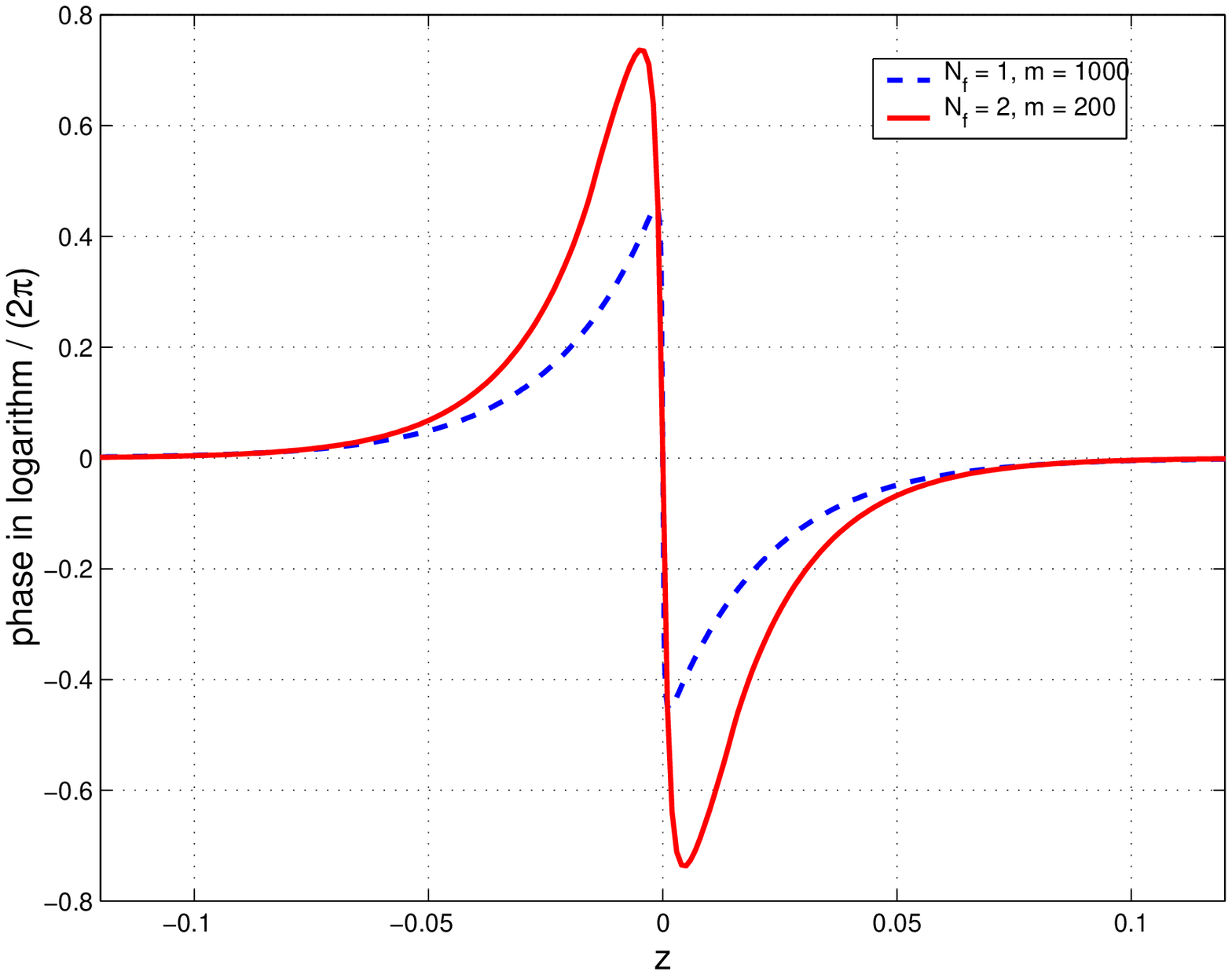}
}
\caption{}
{\footnotesize \noindent \textbf Plot of the phase of the 
logarithm of the TVY superpotential (see Eq.~(\ref{TVY})), in
units of $2 \pi$, as a function of the spatial coordinate, $z$,
for the TVY model with $N_c=5$ and $N_f=1$, $m=1000$ (dashed) and
$N_f=2$, $m=200$ (solid).
}
\label{fig2}
\end{figure}
Also, a more thorough study of the $N_f/N_c < 1/2$ 
case~\cite{Decar01} leads us to claim that, for $N_f>1$, these 
domain walls will cross the logarithmic branch and therefore we
are uncertain of their physical meaning. We illustrate this point 
in Fig.~2, where we plot the phase of the logarithm that appears in 
the second equation of Eqs~(\ref{TVY}) as a function of the spatial 
coordinate, $z$, for a TVY model with $N_c=5$. As we can see, for 
$N_f=1$ and a large enough mass, the total variation of the phase is 
never bigger than $2 \pi$. However, for $N_f=2$ and $m=200$ the
overall variation of the phase is already bigger than $2 \pi$, i.e. 
the fields are crossing the branch when going from one vacuum to the 
other. We can give an estimate of why this is the case. Given the 
boundary conditions we are setting for our field profiles, it is
clear that, at the centre of the wall (i.e. $z=0$)
\begin{equation}
\frac{\beta_0}{2\pi}  =  \frac{k}{2 N_c} , \;\; \quad
\frac{\alpha_{i0}}{2 \pi} = \frac{k}{2 N_c} - \frac{\omega_i}{2}
\;, 
 \label{incond} 
\end{equation}
where, as usual, $k=\sum_i \omega_i$. Note that, by definition, the
phase of the logarithm (given by $(N_c-N_f) \beta + \sum_i
\alpha_i$, modulo $2\pi$) cancels here. It immediately follows that 
the evolution of the phases of the fields from $z=0$ towards the vacuum, 
when $m$ is large, follows a very precise pattern\footnote{This point 
is actually straightforward to realize once the large mass regime has
been explained, see Eq.~(\ref{SeqM}).}: first, the phases
of the matter condensates, $\alpha_i$ change, more or less at the 
edge of the centre of the domain wall (defined by $z=\pm 1/m$); after 
that, the gaugino condensate phase, $\beta$, starts to change as well, 
to compensate the previous changes in such a way that the phase of the
logarithm cancels again at the vacuum. This can be formulated by
saying that, when $z=1/m$, the different phases are given by
\begin{equation}
\frac{\beta_1}{2\pi} =  \frac{k}{2 N_c} , \;\; \quad
 \frac{\alpha_{i1}}{2 \pi}  = \frac{k}{2 N_c} - \omega_i \nonumber    
\;.     \label{midcond} 
\end{equation}
This is the point at which the phase of the logarithm will acquire its 
biggest absolute value, before it starts decreasing again. This is
given by the changes produced both at $\pm 1/m$, i.e.
\begin{equation}
 \frac{|\Delta ({\rm phase \; log})|}{2 \pi} \leq 2 \frac{|k|}{2} 
= |k|  \;\;.  
\end{equation}
Therefore, in order to have a change of the phase less or equal to
$2 \pi$, $|k| =1$ (remember that k must be an integer).

In practice it means that, for equal windings, only models with
$N_f=1$ will give domain walls profiles with no crossing of the
branch. This implies a series of conceptual problems which are beyond 
the scope of this talk, but which should certainly be addressed at 
some point. 
     
Now that the main result has been presented, let us try to understand
why the different situations arise. In order to do that, we consider
the cases where this model could be described by just one of the two
condensates, i.e. either by the gaugino or by the matter condensate.
This corresponds to the limits of very small and very big mass, 
respectively.

$\bullet$ For small masses (i.e. $m \ll \Lambda$) one can integrate
out the gaugino condensate $S$ and describe the theory in terms of the
matter condensate $Q\bar{Q}$ only. The resulting model accounts for 
the results we obtained for $N_f=1$ \cite{Decar99} and also for the 
upper branch of the $N_f=N_c-1$ case, studied by Smilga et al 
\cite{Smilg97}.
     
One can easily check that the second, lower branch present in this 
latter case corresponds to $S \sim 0$. If we have a look at the 
constraint equation (\ref{const_TVY_1}) evaluated at $R_0 \sim0$
the result is
\begin{equation}
\rho_0 = -\frac{N_c}{N_f} \cos \left(\pi \frac{N_f}{N_c} \right) \;\;.
\label{const_low}
\end{equation}
The immediate conclusion one can draw is that there will only exist 
domain wall solutions with $\rho_0>0$ if and only if $\frac{N_f}{N_c}
> 1/2$. It has been argued in Ref.~\cite{Kaplu99} that the existence
of this lower branch is directly connected to the fact that the
K\"ahler metric for the $S$ field is, in this case, singular. In that
case, this lower branch would be nothing but an artifact due to the
choice of metric. It should also be mentioned here that, according to
Ref.~\cite{Binos00}, the total number of BPS-saturated domain walls 
for a given value of $N_f/N_c$ can only change by a multiple of two, 
when $m$ is continuously varied, and that is the reason why the upper 
and lower branches end up collapsing into a non-BPS domain wall above 
$m_*$. However this does not explain why they have to collapse at all.
   
$\bullet$ For large masses (i.e. $m \gg \Lambda$) one can integrate 
out the matter condensate ($Q\bar{Q}$) and our model will be described
by the gaugino condensate only (i.e. we recover SUSY gluodynamics). 
In order to do that, it is worth noticing that, in this large mass 
regime, $M \sim S$, and we can do the following identifications
\begin{eqnarray}
\rho(z) e^{i \alpha(z)} & = & R(z) e^{i\beta(z)} \;, \;\;\; z \ll -1/m 
\nonumber \\
\label{SeqM} \\
\rho(z) e^{i \alpha(z)} & = & R(z) e^{i(\beta(z)-2 \pi)} \;\;, \;
z \gg 1/m \nonumber \;\;.
\end{eqnarray}
That is, to the left of the centre of the domain wall, both
condensates behave in the same way whereas, to the right of it,
their phase difference is $2 \pi$. This is illustrated in Fig.~3,
where we plot the phase difference between the two condensates (in
units of $2\pi$) as a function of the spatial coordinate, for several
values of the mass parameter $m$.
\begin{figure}
\centerline{  
\includegraphics[width=7cm]{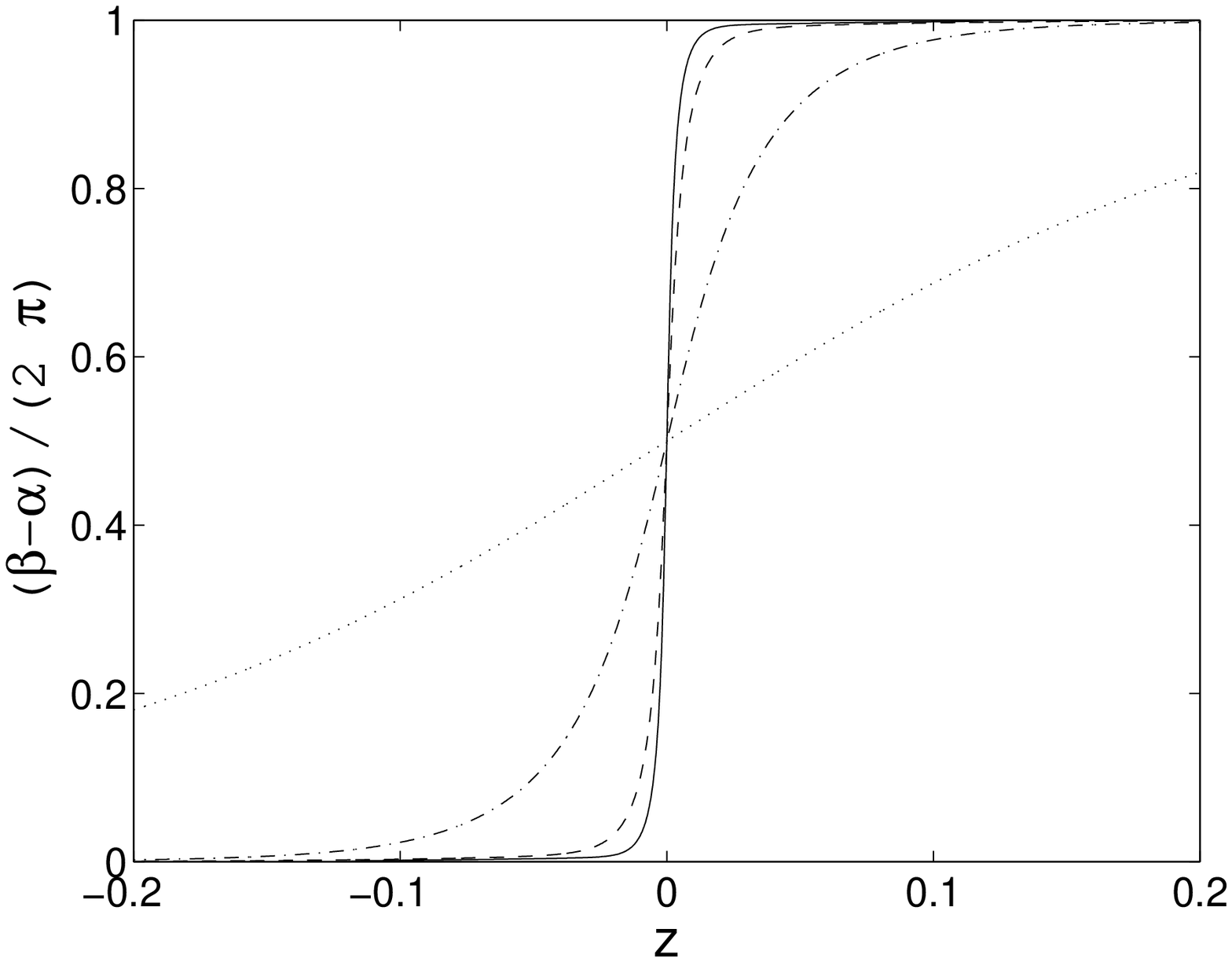}
}
\caption{}
{\footnotesize \noindent \textbf Plot of the phase difference 
between condensates, $(\beta-\alpha)$ (in units of $2 \pi$), as a 
function of the spatial coordinate, $z$, for $N_c=3$, $N_f=1$ and 
different values of the mass of the matter condensate: $m=2$ (dotted), 
$m=20$ (dash-dotted), $m=100$ (dashed), $m=200$ (solid).
}
\label{fig3}
\end{figure}
One can also analyze the constraint equation (\ref{const_TVY_1}) in
this large mass limit, which will be given in terms of the 
modulus of the condensate $R_0$ as
\begin{equation}
R_0 (1- \ln(R_0)) = \cos \left( \pi \frac{N_f}{N_c} \right) \;\;.
\label{const_high}
\end{equation}
A quick glance at this equation tells us that, in order to have
$R_0<1$, which corresponds to finite-energy, well-defined domain
walls, we must be in the case where $\frac{N_f}{N_c} < 1/2$. In
other words, it does not seem possible to reach the $m \rightarrow
\infty$ limit, continuously from small $m$, in a model where 
$\frac{N_f}{N_c} > 1/2$. The fact that, in that case, there are no    
BPS-saturated domain walls at large $m$, whereas we had two branches
of solutions at small $M$ explains why, at some point, those must
have annihilated each other at $m_*$. 
     
On the other hand, Eq.~(\ref{const_high}) is telling us that there 
is an analytic limit $m \rightarrow \infty$ limit that we can
construct when $N_f/N_c < 1/2$; using Eq.~(\ref{SeqM}) we can write 
down two BPS equations for $S$, one for the right hand side of the 
domain wall and another one for the left hand side. The solution is 
given by the thick line in Fig.~1 and, as we can see, coincides very 
well with what we would expect for the large $m$ limit by looking at 
the finite mass results.
     
\subsection{Different windings}
     
Now that we have presented and explained most of the results in the 
literature, let us step onto new ground. As mentioned above, up
to now the standard procedure was to consider all the matter fields 
transforming in the same way when going from one vacuum of the theory
to another. We shall now break this degeneracy and assign different
winding numbers to the different matter fields. At this stage we 
present results concerning a particular example, that of $N_c=3$ and 
$N_f=2$, while a more detailed and general analysis will be 
presented elsewhere \cite{Decar01}.

According to the results of the previous subsection, when $N_c=3$ and
$N_f=2$, then $N_f/N_c>1/2$ and we should expect two branches of 
BPS-saturated domain walls that annihilate each other at $m_*$ (see
Ref.~\cite{Smilg97}). And that is indeed what happens when 
$\omega_1=\omega_2=1$. However, if we consider a different choice for 
the windings, i.e. $\omega_1=1$, $\omega_2=0$, then it turns out that 
one can construct BPS-saturated domain walls for {\em any} value of
the mass $m$. The profiles of these fields can be seen in Fig.~4, 
where we plot the Argand diagram for $S$, $M_1$ and $M_2$ with masses 
$m_1=m_2=250$.
\begin{figure}
\centerline{  
\includegraphics[width=7cm]{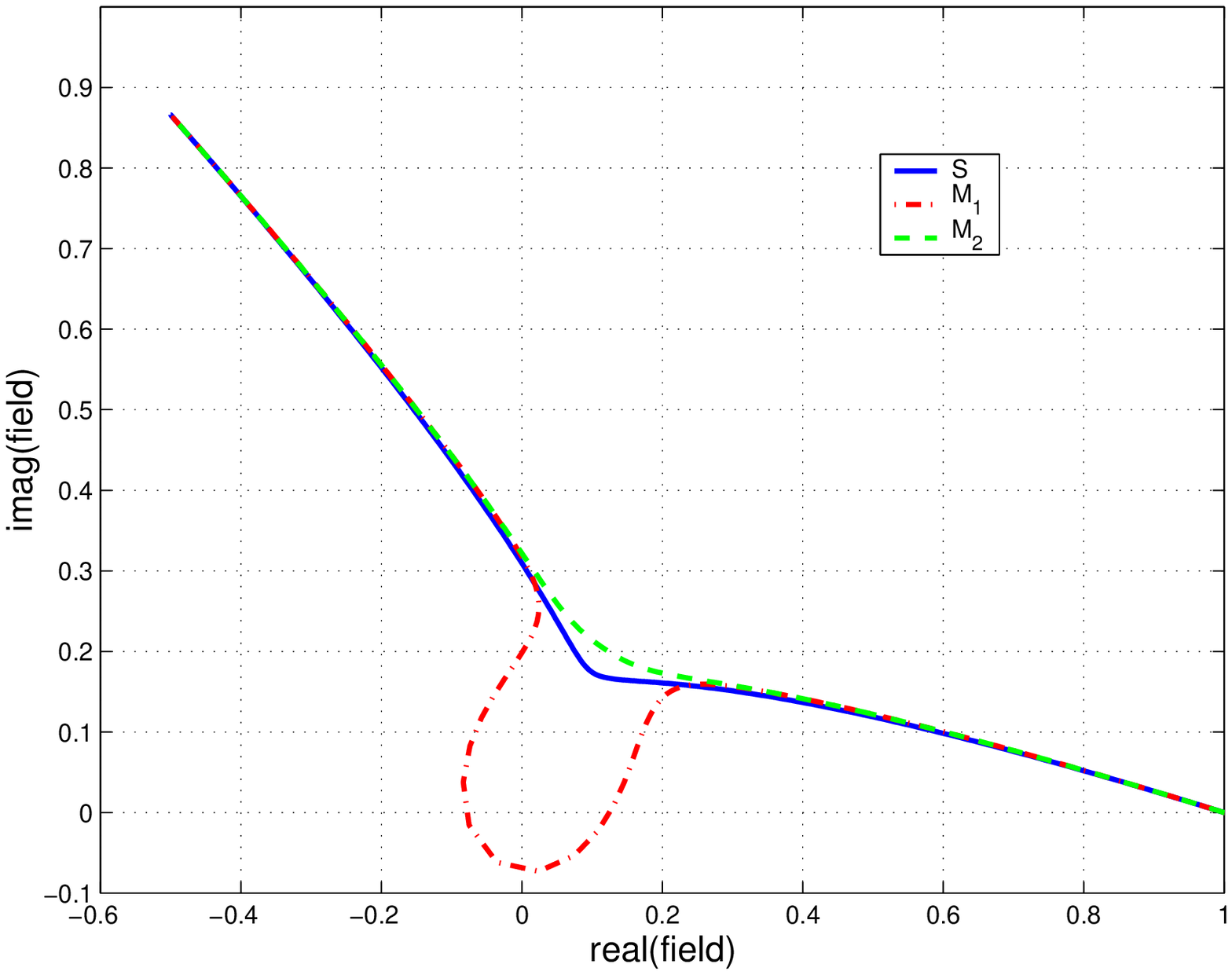}
}
\caption{}
{\footnotesize \noindent \textbf Argand diagram of the profiles
of the different condensates for $N_c=3$, $N_f=2$. Here $\omega_1=1$,
$\omega_2=0$, $m_1=m_2=250$. The gaugino condensate, $S$, is given
by the solid line while $M_1$ and $M_2$ correspond to the dot-dashed 
and dashed lines respectively.
     }
\label{fig4}
\end{figure}
There it can be seen that the matter condensate with $\omega_1=1$ 
follows a different path than the one with no winding, which simply 
mimics the gaugino condensate, $S$; and it is exactly the same path 
described by the matter field for $N_c=3$, $N_f=1$. Actually it is 
possible to increase the value of $m_2$ while keeping $m_1$ finite 
and constant, in order to integrate out the second condensate and 
recover the results we obtained for $N_c=3$, $N_f=1$. The key point 
to perform this exercise is to notice that the relevant ratio that
classifies the solutions is not $N_f/N_c$ but $k/N_c$ where, as 
defined before, $k=\sum_i \omega_i$. In the previous case, $k=N_f$, 
whereas now we have $N_f=2$ with $k=1$. 
     
Therefore it should be in principle possible to construct BPS-saturated 
domain walls for {\em any} value of $N_f$, $N_c$, just by choosing the 
windings of the matter fields in such a way that their sum, $k$, over the 
number of colours, $N_c$, is less than $1/2$. However it must be pointed 
out once again that, only when $|k| \leq 1$, the paths described by the 
different fields between two vacua do not cross the logarithmic branch.
Those are the only physically meaningful domain walls that we know, so 
far, how to construct.    

\section{Conclusions}

Let us summarize first the results obtained, before discussing what would
be interesting to pursue in order to improve our understanding of these
objects. We have looked in detail at domain walls in SUSY-QCD, as a way
of understanding SUSY gluodynamics, the nearest relative to ordinary QCD.
Using the TVY effective Lagrangian, we have constructed BPS-saturated
domain walls for certain values of $N_f$, the number of matter condensates,
and $m$, the mass of those condensates. The results so far obtained
can be summarized as follows:
\begin{itemize}
\item All matter fields transforming in the same way (equal windings)
 
   \begin{itemize}
    \item If $N_f/N_c< 1/2$ there are BPS-saturated domain walls for 
{\em any} value of $m$.
   
   \item If $N_f/N_c \geq 1/2$ there are BPS-saturated domain walls up
    to $m_*$.
   
   \item The logarithmic branch of the scalar potential is not crossed 
   only when $N_f=1$.
  \end{itemize} 

\item Non-degenerate flavours (different windings): any choice of $\omega_i$
such that $k=\sum_i \omega_i=1$ gives BPS-saturated solutions for {\em any}
($N_f$,$N_c$) and $m$ with no crossing of the branch.

\item The analysis of the constraint ${\rm Im} (e^{i\gamma} {\cal W})=
{\rm const}$ at the origin ($z=0$) is the key to understand the results.
\end{itemize}     
     
So, at least, we have now a criterion to construct well-behaved 
BPS-saturated domain walls but, of course, this project is far from
over. The next issue that should be addressed is that of the branches 
associated with the logarithm in the potential. As mentioned at the end
of section~2, several suggestions have been put forward but none has 
so far provided us with a satisfactory answer.

Another issue which is rather controversial is the dependence of these 
results on the choice of K\"ahler metric. In fact it has been widely 
claimed in the literature that the origin of the lower branch found in 
Refs~\cite{Smilg97} is a K\"ahler metric which blows up near the
centre of those solutions. This is not a problem that one encounters 
in theories with higher supersymmetries, where the K\"ahler function 
is totally determined by the symmetries of the theory. Therefore an 
obvious way of trying to understand this dependence is through the 
connection of the $N=1$  SUSY contructions here presented with those 
attempted with higher supersymmetries (for example, those of
Refs.~\cite{Kaplu99,Dorey00}). 
     
Finally it would be desirable to connect the results obtained here 
with constructions of BPS-saturated domain walls in the large $N_c$
limit in the context of SUSY gluodynamics, which have been performed 
in Ref.~\cite{Dvali99}. As mentioned in the introduction, these
objects could play a very important role in the D-brane description of 
SUSY-QCD.

     
\end{document}